# New Algorithm to Investigate Neural Networks[1]

Bernd A. Berg[2,3,4]

**Abstract**

Random cost simulations were introduced as a method to investigate optimization problems in systems with conflicting constraints. Here I study the approach in connection with the training of a feed-forward multilayer perceptron, as used in high energy physics applications. It is suggested to use random cost simulations for generating a set of selected configurations. On each of those final minimization may then be performed by a standard algorithm. For the training example at hand many almost degenerate local minima are thus found. Some effort is spent to discuss whether they lead to equivalent classifications of the data.

[1]This research was partially funded by the Department of Energy under contract DE-FG05-87ER40319.
[2]Department of Physics, The Florida State University, Tallahassee, FL 32306, USA.
[3]Supercomputer Computations Research Institute, Tallahassee, FL 32306, USA.
[4]E-mail: berg@hep.fsu.edu

# 1 Introduction

Recently there has been some interest [1, 2, 3, 4, 5] in Monte Carlo (MC) sampling from Broad Energy Distributions (BED). The basic idea is about twenty years old and was first introduced under the name Umbrella Sampling [6]. The increased interest in related methods began with the success of Multicanonical Sampling [1] in the study of first order phase transitions. The name *multicanonical* emphasizes the possibility of obtaining from one sample canonical expectation values over a temperature range. Soon a wide range of applications was realized. In particular it was stressed that the algorithmic ergodicity becomes enhanced by sampling with BED. This has lead to new perspectives concerning numerical investigations of systems with conflicting constraints, like for instance spin glasses [2, 4], proteins [7] or the traveling salesman problem [8].

A complication of these approaches is that they sample with weight factors $w(E)$ which are a-priori unknown functions of the energy $E$. It is part of the algorithm's purpose to converge to a suitable approximation, which then allows to estimate the spectral density $\rho(E)$. In practice complications emerge which are unknown for canonical MC simulations, where the correct weights are given by the Boltzmann factor $w_B(E) = \exp(-\beta E)$.

The Random Cost (RC) method [9] samples a BED without the need of tedious recursions towards appropriate weight factors. This is achieved by employing simple master equations to enforce a random walk in a given cost function, for instance in the energy of a statistical mechanics system. The price paid is that one does not sample anymore with weights which depend only on the energy (*i.e.* the cost function). Consequently the ability to construct canonical expectations values is lost. This disadvantage is presumably of minor importance in applications to hard optimization problems, where one is mainly interested in an overview of the minima of the system and less in its statistical mechanics. RC may then compete with approaches like simulated annealing [10] or genetic algorithms [11].

In ref.[9] the RC method was illustrated for an artificially simple cost function. Since then, no new experience was reported. One reason, as we shall see, is that implementing the method in more realistic situations is not entirely straightforward. There is a large amount of innovative freedom in setting up the random walk master equations. Realistic applications require to make some decision and wrong ones render the algorithm ineffective.

In the present paper I focus on applying the basic ideas to the training of Neural Networks (NN). In high energy physics NN constitute powerful nonlinear extensions of conventional data analysis methods, see [12, 13, 14] and references therein. In the context of this paper the purpose of the NN is to illustrate (a) how the RC method works and (b) how it may lead to interesting new physical insight. The training of a feed-forward two-layer perceptron to search for top quark production in "all-jet" channels [16] is considered. The RC simulations yield a large number of local minima, which are well-separated in parameter space. This allows to address relevant questions like:

(i) Is one global minimum dominating or are there many almost degenerate minima?

In case of many almost degenerate minima:

(ii) What is their distribution in parameter space?

(iii) Do different minima lead to the same, or at least to similar, classifications of the data into events and background?

The NN and the training data are described in the next section. In section 3 the RC method is outlined in some details. Section 4 is devoted to numerical results and their interpretation. On their basis, I conclude that RC is a promising method in the context of exploring NN minima. One may hope for considerable further improvements by exploiting the innovative freedoms of the method more efficiently. Summary and conclusions are given in section 5.

## 2 The Training Example

We shall consider the training of a feed-forward two-layer perceptron for $t\bar{t}$ detection through $b$-quark tagging with soft muons [16]. The network function is defined by

$$Y_k = g\left[\sum_{i=1}^{m} \omega_i^2\, g\left(\sum_{j=1}^{n} \omega_{ij}^1\, d_{jk} + \theta_j^1\right) + \theta^2\right] \quad \text{where} \quad g(x) = \frac{1}{1+\exp\left(-2x\right)}\,. \tag{1}$$

Here $d_{jk}$, $(k = 1, ..., N_d)$ are experimental data and $\omega_{ij}^2$, $\theta^2$, $\omega_{ij}^1$, $\theta_i^1$ are the parameters of this network. In our example $m = 5$ and $n = 4$. Hence, there are $5\,\omega_i^2$, $1\,\theta^2$, $20\,\omega_{ij}^1$ and $5\,\theta_i^1$. This leaves us with 31 parameters which, generically, will now be denoted by $x = (x_j)$, $(j = 1, ..., 31)$. The aim of a training program is to minimize the mean square error

$$E_2 = \frac{1}{N_d} \sum_{k=1}^{N_d} \left(Y_k - \theta(N_b - k)\right)^2 \,. \tag{2}$$

The function $Y_k$ itself is not binary, but has the useful property that (under certain conditions) it can be interpreted as a Bayesian a posteriori probability [15]. We shall use $N_d = 5000$ data $d_{jk}$ to train the network. For $k = 1, ..., 2500$ they are from the D0 50K sample [16, 17], and used to train the NN for background. This is achieved by choosing $N_b = 2500.5$, *i.e.* $\theta(N_b - k) = 1$. (The likelihood that a data point describes an event is less than 1/1000 for these data.) For $k = 2501, ..., 5000$ the data are MC generated events from the ISA180_ALL.HBOOK sample. Each data point is a standard AllJets 4-tuple [16]

$$d_j = (C, APL, NJ1/10, HT3/500),\ (j = 1, 2, 3, 4)\,.$$

The symbols stay for the following global event quantities: $C$ = centrality, $APL$ = aplanarity, $NJ1$ = average jet count, and $HT3$ = sum of jet $E_T$ excluding the first two jets.

## 3 Random Cost Simulations

We are interested in finding many (local) minima of a function

$$f = f(x) \ \text{ where } \ x = (x_j) = (x_1, ..., x_n)\,,$$

and in that process possibly its global minimum. In hard optimization problems (problems with conflicting constraints) it happens that one has to overcome barriers (local increases) of the function $f(x)$ before convergence into globally interesting minima is achieved. The purpose of the RC method is to overcome such barriers through a stochastic process. In essence the method is described by the following four steps.

(i) Generate randomly a set of update proposal for the argument: $\{\triangle x_{(k)}\}$. (Here we distinguish different function arguments by subscripts in parenthesis, like $x_{(k)}$, whereas components of the argument are singled out through subscripts without parenthesis, like $x_j$.)

(ii) Calculate the function changes $\triangle f_{(k)} = f(x + \triangle x_{(k)}) - f(x)$.

(iii) Divide the update proposals into three subsets. First, $\{\triangle x^+_{(k)}\}$ and $\{\triangle x^-_{(k')}\}$ are defined such that $\triangle f^+_{(k)} > \triangle f_{\min}$ and $\triangle f^-_{(k')} < -\triangle f_{\min}$ holds for the corresponding function changes. Here $\triangle f_{\min} > 0$ is some (small) cut-off. All update proposals with $|\triangle f^0_{(k'')}| \leq \triangle f_{\min}$ form the third set, $\{\triangle x^0_{(k'')}\}$.

(iv) When both, the $\{\triangle x^+_{(k)}\}$ and the $\{\triangle x^-_{(k')}\}$ set, are non-empty: Updates from these sets are chosen according to a probabilistic law which enforces a random walk in the function value $f$. (In case that one of these sets is empty, violations may be allowed.)

In this paper the set of update proposal is defined as follows. Each component $x_j$ is restricted to the same range $|x_j| \leq x_{\max}$. Allowed are updates in steps of $\triangle x_{ij}$ with

$$\triangle x_{ij} = \text{sign}\,(i)\; 2^{-1-|i|}\, x_{\max}\,, \quad \text{with} \quad i = \pm 1, \pm 2, ..., \pm i_{\max}\,.$$

The subscript $i$ labels the stepsize and sign, whereas $j$ picks a component of $x$. The updates are thus confined to a grid. The minimum grid length $\triangle x_{min}$ is determined by the choice of $i_{\max}$. I like to emphasize that my choice of update proposals is neither unique nor claimed to be particularly efficient. The method allows for all kind of choices and presently it is unclear by which criteria efficient ones may be singled out.

In [9] the entire $\triangle f_{ij}$ array was calculated for each RC update. For the present, more realistic, cost function the computational effort becomes then considerable. Fortunately, it turns out to be rather straightforward to invent modified updating procedures which are far less CPU time intensive. The simulations of the next section relies on the following one:

Elements of the $\triangle x_{ij}$ array are picked at random [18] and the corresponding $\triangle f_{ij}$ elements are calculated. As soon as $\triangle f_{ij} > \triangle f_{\min}$ and $\triangle f_{i'j'} < -\triangle f_{\min}$ elements are found, the RC update is performed. Let us first assume that this happens before the entire array $\triangle x_{ij}$ is exhausted. Then either for $\triangle f_{ij} > \triangle f_{\min}$ or for $\triangle f_{i'j'} < -\triangle f_{\min}$ there will be precisely one proposal. ¿From the other set, one element is picked at random. As the elements were already picked at random, it is sufficient to to chose the last element. This means, we have two definite updating proposals

$$\triangle x_j^- \text{ corresponding to } \triangle f^- < -\triangle f_{\min}$$

and

$$\triangle x_{j'}^+ \text{ corresponding to } \triangle f^+ > \triangle f_{\min}\,.$$

The RC equation is then simply

$$p^- \triangle f^- = p^+ \triangle f^+ \tag{3}$$

This equation is easily solved for, say,

$$p^- = \frac{\triangle f^+}{\triangle f^- + \triangle f^+}\,. \tag{4}$$

A random number $x_r$, uniformly distributed in the range $0 < x_r \leq 1$, is then chosen. For $x_r \leq p^-$ the $\triangle x_j^-$ update is accepted, otherwise the $\triangle x_{j'}^+$ update.

When the entire set $\triangle x_{ij}$ leads only to updates with either $\triangle f_{ij} > -\triangle f_{\min}$ or $\triangle f_{ij} < \triangle f_{\min}$, we have found a local minimum or maximum. To be precise, we have found a local minimum or maximum within the precision imposed by the cut-off choice $\triangle f_{\min}$. In its present implementation the simulation continues by accepting the last proposed $\triangle x_{ij}$. The function values $f$ will perform a random walk between thus defined local minima and maxima. If $\triangle f$ is a typical stepsize and $f_{\max} - f_{\min}$ a typical distance between a local maximum and a local minimum, the simulation will need of the order $|f_{\max} - f_{\min}|^2/|\triangle f|^2$ steps to get from one side to the other. Here $|\triangle f|$ is bounded from below by $\triangle f_{\min}$. Related to this, a too small choice of $\triangle f_{\min}$ renders the simulation inefficient. Instead of aiming at reaching local minima with high precision, it is here suggested to record the time series for a reasonable choice of $\triangle f_{\min}$. Many independent regions of configuration space are then reached. Independent minima of the time series are subsequently taken as starting points for one of the conventional [19] downward minimization algorithms.

One may further restrict the RC simulations by imposing additional bounds. For instance one may reject all updates which lead to a function value larger than an imposed maximum $f_{\max} > 0$. Or one may reject all updates with $\triangle f > \triangle f_{\max}$ where, of course, $\triangle f_{\max} \gg \triangle f_{\min} > 0$. Some experience with such bounds is reported in the next section. As a general rule, I like to suggest that upper bounds on the function value should be imposed in a stochastic way by modifying the RC equation (3) in favor of one direction.

# 4 Numerical Results

Results from RC simulations of the NN error function (2) are now reported. Algorithmic performance and applications of physical relevance are treated in different subsections.

## 4.1 Algorithmic performance

For the parameters, discussed in the previous section, the following choices are made:

$$\triangle f_{\min} = 10^{-4},\ \triangle f_{\max} = 0.1\,,$$

and no upper bound $f_{\max}$. Further

$$|x_j| \le x_{\max} = 2.5 \ \text{ and } \ i_{\max} = 12 |x_j| \le x_{\max} = 10 \ \text{ and } \ i_{\max} = 14$$

were tested. In each case $\triangle x_{\min} = 5/2^{14} \approx 0.00061$. Distribution functions are defined by

$$F(E_2) = \int_0^{E_2} \rho(E_2')\, dE_2' ,$$

where $\rho(E_2)$ is the corresponding probability density. In practice estimators are obtained by simply sorting [19] the sampled values of $E_2$. To plot distribution functions, instead of histograms of the probability density $\rho(E_2)$, has the advantage that one needs not to worry about an appropriate bin size. Figure 1 compares (for two $x_j$ ranges) the distribution functions from RC simulations versus those from random sampling (RS). Here a RS configuration is defined by choosing for each parameter a random number, uniformly distributed in the allowed range.

The reader should focus on the behavior of $F(E_2)$ for small $E_2$ values. The RC distributions show a sharp increase: For $|x_j| \leq 2.5$ about 20% of the configurations are generated in the range $E_2 \leq 0.2$. For $|x_j| \leq 10$ this values is even up to more than 30% , implying that this is the preferable RC parameter choice. In contrast to RC simulations, RS generates almost no configurations in the $E_2 \leq 0.2$ range. It is amazing to note that for RS the parameter range $|x_j| \leq 2.5$ is preferable. For $|x_j| \leq 10$ most RS configurations exhibit $E_2$ values very close to 1/2.

Concerning the RC results, it should be noticed that the distribution functions are not straight lines due to the fact that the magnitude of a typical change $\triangle E_2$ ($E_2$ is the function $f$ of section 3) depends on $E_2$. In the neighbourhood of local minima (in the sense of the algorithm) $\triangle E_2$ proposals become small and the algorithm spends more time there. Of course, configurations related by small $\triangle E_2$ changes are strongly correlated.

To find out how many independent minima are generated, I depict in figure 2 the RC time series for the better parameter choice ($|x_j| \leq 10$). Altogether 100,000 RC updates (changes of a single parameter) are performed. For each 1000 updates the minimum and maximum values reached are plotted and, in order of their occurance in the time series, connected by straight lines to guide the eyes. Autocorrelation are clearly visible, but at the same time it becomes clear that a large number of independent minima (certainly $> 20$) are created.

Independent minima may be singled out by requiring that the time series went over some cut-off barrier $E_2^c$ between subsequently recorded minima. From figure 2 as well as from the nature of the problem it is clear that $E_2^c = 0.4$ is a reasonably high choice. The lowest twenty minima left over then are depicted in figure 3 together with the lowest twenty minima obtained by creating $10^6$ RS configurations. On a DEC 3000 Alpha 600 workstation the CPU time needed for the 100,000 RC updates was 12.2 hours and the CPU time to create $10^6$ RS configurations was about 13 hours. It is obvious that RC easily outperforms RS also when autocorrelations are taken into account.

Ideal efficiency of RC would be expected when energy barriers populate the region in-between the minima reached by RS and those reached by RC. This is due to the feature that RC climbs as enthusiastically uphill as downhill. It works by suppressing the statistical weight of configurations in-between extrema. In our example there are no strong indications of such barriers. The better performance of RC seems to be entirely due to the fact that it samples the rare configurations with low (and high) $E_2$ far better than RS. In this sense the present case is too simple for RC. It remains to be explored whether NN with actual barriers between the RS region and the RC minima do exist.

A (primitive) steepest gradient minimization program was applied to the RC as well as to the RS configurations whose $E_2$ values are shown in figure 3. The purpose is to converge to the local minimum closest to the starting configuration. After this minimization the average $\overline{E_2}$ and best $E_{2,\min}$ values were

$$\overline{E}_2 = 0.11831 \pm 0.00025, \ \ E_{2,\min} = 0.11586 \ \ \text{for RC}$$

and

$$\overline{E}_2 = 0.11896 \pm 0.00013, \ \ E_{2,\min} = 0.11788 \ \ \text{for RS.}$$

The configurations thus found are called RC (or RS) minima in the following. Although the difference in the mean value $\overline{E}_2$ is not very dramatic, it is notable that the first seven RC minima are all lower than the best RS minimum.

Default settings of JETNET [12] return the value $E_2 = 0.11722$ [17]. This would put it at position 5 in my set of RC minima. Running my minimization program on the configurations produced by JETNET reduces this value further to $E_2 = 0.11620$. This is the second best of all my solutions and obtained far more CPU time efficient than the others. The point of RC is clearly not to save CPU time. Instead the purpose it to provide a simple method which allows to explore relatively hassle free relevant regions of the configurations space. Nowadays, it is normally a minor problem to find a fast workstation for a few days of MC simulations. To program a complicated approach could be the real stumbling block. The aim of a RC simulations is to gain increased confidence, that relevant regions of configuration space have not been overlooked. RS serves this purpose far less well, because the entropy of the interesting regions tends to be very small. If, in addition, energy barriers separate relevant minima from the high entropy region, RS with subsequent minimization may not get to them at all. Adding Gaussian noise to minimization certainly helps, but the entropy preference of such noise is the same as that of RS.

RC greatly suppresses the high entropy regions while, at the same time, being able to climb up and down. Simulated annealing achieves the same purpose by varying the temperature. (The function $E_2$ is then interpreted as the energy of a statistical mechanics system. It should be noted that RS corresponds to infinite temperature $\beta = 0$.) Here an advantage of RC seems to be that it needs less detailed considerations. Parameter choices like $\triangle f_{\rm min}$ or $x_{\rm max}$ are needed in both approaches. RC is then ready for a long run, as eqns. (3) automatically ensure a broad distribution (figure 1). In simulated annealing one has to worry about a scheme for lowering (and possibly rising again) the temperature. In many applications one may be unwilling to spend the work it takes to tune an annealing scheme. Such a scheme is necessary, because statistical mechanics distributions are narrow at any fixed temperature.

If desired RC allows some tuning too. In particular, as we are interested in minima, one may like to restrict the sampling region by introducing an upper bound $f_{\rm max}$. This should be done in a smooth way. Figure 4 compares the $F(E_2)$ distribution functions for a sharp versus a smooth upper bound $f_{\max} = 0.3$. The smooth bound is achieved by doubling the $p^-$ value of equation (4) for $E_2 > 0.3$. It is clear that the simulation with the sharp upper bound is the worse: It spends a large amount of CPU time on the immediate neighborhood of $E_2 = 0.3$, because the updating stepsize $\triangle f$ approaches there $\triangle f_{\min}$. The simulation with the smooth upper bound moves far more freely in the $E_2 = 0.3$ neighborhood. Consequently, it spends less CPU time there and still reaches distant configurations faster. It should be noted that no major improvement over the simulation without upper bound was achieved. For the smooth bound the minima yield $\overline{E}_2 = 0.11772 \pm 0.00012$ and $E_{2,\min} = 0.11680$.

In difficult situations it may be worthwhile to try RC as one of various approaches. Each method, simulated annealing [10], multicanonical annealing [8], genetic algorithms [11] or RC has its own specific way to explore configurations space. Which method wins is most likely problem dependent. Presently there are no a-priori criteria at hand to choose one method over the others. Not spending too much of your own time may well favour RC.

## 4.2 Physical applications

The physical purpose of finding many minima is to increase confidence in classifications proposed by a NN. It is after all some kind of black box. At the first look differences between the twenty RC minima are rather small. To make the point, let us consider the RC minima with lowest and highest $E_2$. In figure 5 distribution functions $F(Y_k)$, with $Y_k$ defined by equation (1), for the event and background training data of these solutions ($E_2 = 0.11586$ and $E_2 = 0.11984$) are plotted. Events are the upper curves and background are the lower curves.

It is seen that the smaller $E_2$ value comes from the fact that this solution concentrates the background more efficiently into the $Y_k \to 1$ limit. The other solution concentrates events more efficiently into the $Y_k \to 0$ limit. Apparently, the price paid is that also some of the background events get placed into this limit, as is more clearly seen from the inlay.

Altogether one tends to conclude that the classifications are almost equivalent, the main difference being that the entire curve is shifted with slight distortions of the shape. However, one has to make sure that there is not internal re-ordering, *i.e.* identical data classified far apart in different distributions of similar shape. To address this and other questions, it is convenient to introduce some norms. Let $X = (X_1, ..., X_n)$ with $0 \le X_i \le 1$, we define:

$$||X||_1 = \frac{1}{n}\sqrt{\sum_{i=1}^{n}(X_i)^2}\,,\ ||X||_2 = \max\{|X_i|,\, i=1,...,n\} \ \text{ and } \ ||X||_3 = \frac{1}{n}\sum_{i=1}^{n}|X_i|\,. \tag{5}$$

Relying on these norms various average distances were calculated and are reported in table 1.

Let us first address the distances in parameter space. The parameters of the twenty RC minima are denoted by

$$x^{(s)}_{\text{min}} = (x^{(s)}_1, ..., x^{(s)}_{31}) \ \text{ where } \ s = 1, ..., n_s$$

and $n_s = 20$ for the RC results. The second column of table 1 gives

$$\langle ||x^{(s)}_{\text{min}} - x^{(s)}_{\text{select}}|| \rangle = \frac{1}{n_s}\sum_{s=1}^{n_s} ||x^{(s)}_{\text{min}} - x^{(s)}_{\text{select}}||\,,$$

the average distance of the minima $x^{(s)}_{\text{min}}$ away from the starting values $x^{(s)}_{\text{select}}$, which are selected from the RC simulation runs before local minimization is applied. When calculating the norms each $x^{(s)}$ component is first rescaled from the $[-10, 10]$ range into $[-0.5, 0.5]$, because a range of length one is used in the definition (5). The error bars in parenthesis apply to the last digits and correspond to our statistics of twenty solutions. Column two of table 1 should be compared with column four, where the (up to the given digits) exact average distance of 32 component independent random vectors $x^{(s)}_{\text{ran}}$ and $x^{(t)}_{\text{ran}}$ is written down. For these vectors each component is an uniformly distributed random number in the range [0,1). As expected, the found minima $x^{(s)}_{\text{min}}$ are fairly close to the parameters $x^{(s)}_{\text{select}}$ selected from the RC simulation.

Column three collects the average distances

$$\langle ||x^{(s)}_{\text{min}} - x^{(t)}_{\text{min}}|| \rangle = \frac{1}{n_s\,(n_s-1)}\sum_{s=1}^{n_s}\sum_{t=1}^{n_s} ||x^{(s)}_{\text{min}} - x^{(t)}_{\text{min}}||$$

between different minima. There are $19 \cdot 20/2 = 190$ different combination, to which the average values correspond. As only twenty are independent, the error bars are obtained as $\sqrt{\sigma/19}$ from the estimated variances. It is seen that these distances are close to the distances between random vectors. A plot of all parameter values found is given in figure 6 and looks very similar to a plot were uniform random numbers in the $[-10, 10]$ range are drawn for each parameter. It goes beyond the scope of this paper to analyse for correlations.

Let us turn to distances in function space. The vector components are then of the form $Y_k^{(s)}$. For column five and six $s$ and $t$ label again our twenty RC minima and $k = 1, ..., 5000$ labels the training data. In column seven results for a random vector with 5000 components are reported.

For the average distances $||Y_{\rm sort}^{(s)} - Y_{\rm sort}^{(t)}||$ the $Y_k^{(s)}$ and $Y_k^{(t)}$ components are sorted in increasing order before the norms are calculated. These distances are characteristic for the differences seen in plots like figure 5 (but note that now all data are accommodated in one distribution function). For all three norms these distances turn out to be about 5% of the distances $||Y_{\rm ran}^{(s)} - Y_{\rm ran}^{(t)}||$ which one encounters for random vectors of length 5000.

The average distances $||Y^{(s)} - Y^{(t)}||$ are the quantities of physical interest: They relate directly to differences in the classification of our training data. For norm one and norm three the numbers are about two times those of the corresponding sorted vectors, *i.e.* about 10% of the random vector results. However, a few data points behave exceptional. The result for norm two means that the worst average re-ordering amounts to about 58% percent of the function values range ($0 \leq Y_k \leq 1$). As these are averages taken over the 190 possible combinations of our solutions, the re-ordering of certain data with respect to two different network solutions is even worse. Indeed the largest re-ordering encountered is $||Y^{(14)} - Y^{(10)}||_2 = 0.91$, whereas the best of the worst is $||Y^{(17)} - Y^{(4)}||_2 = 0.21$. For the solution $Y^0$ found via JETNET and the best RC minimum we have $||Y^{(1)} - Y^{(0)}||_2 = 0.31$. Finally for the solutions depicted in figure 5 the value is $||Y^{(20)} - Y^{(1)}||_2 = 0.57$. These results show that the (slight) increase of $E_2^{(s)}$ with $s$ for $s = 1, ..., 20$ seems to be irrelevant for the

reordering effect. Instead the internal structure of the solutions should be hold responsible.

The small $\langle ||Y^{(s)} - Y^{(t)}|| \rangle$ averages obtained with the other two distance definitions imply that large re-ordering happens only for a few data points. This is confirmed by plotting the distribution function of the $|Y_k^{(s)} - Y_k^{(t)}|$ average in figure 7. For 98% of the data $\langle |Y_k^{(s)} - Y_k^{(t)}| \rangle$ is less than 0.1. It should be noted that the highest value for $\langle |Y_k^{(s)} - Y_k^{(t)}| \rangle$ is lower than $\langle ||Y_k^{(s)} - Y_k^{(t)}||_2 \rangle$ of the table, because the $k$ values for which the largest value is obtained depends on $s$ and $t$. Of course, it is no problem to identify the individual data points which are subject to large re-ordering. It may be interesting, but is beyond the scope of this paper, to investigate whether they exhibit particular physical characteristics.

To what extent can one now trust a classification proposed by the NN? The worst case scenario combines different solutions in the following way:

$$W_k^n = \max\{Y_k^{(s)} | s = 1, ..., n\} \text{ for events}$$

and

$$W_k^n = \min\{Y_k^{(s)} | s = 1, ..., n\} \text{ for background}.$$

Here a cut off on the maximum allowed $E_2$ value has to be set. A value of the order of a few percent seems to be reasonable. In the situation at hand, the $\triangle E_2$ difference between solutions # 1 and # 20 is about 3.5% . Figure 8 shows what happens when solutions $s = 1, ..., 20$ are successively combined according to the worst case scenario. In the region $0.1 \le W_k^n \le 0.9$ results apparently get stable. However, in the extreme limits (*i.e.* for a small amount of data) crossover effects between classification as event versus background are found. The results suggest that one should not apply this NN in these limits.

# 5 Summary and Conclusions

RC simulations sample ergodically through configuration space, while greatly enhancing (as compared to RS) the likelihood of configurations in the neighbourhood of minima (or maxima). The updating scheme employed in this paper is considerably improved over the version of [9]. Further significant progress in this direction seems to be likely.

A large number of practically independent local minima may be obtained by combining RC simulations with subsequent minimization. Many regions of configuration space are thus covered and barriers between them can be overcome. This increases the confidence that best solutions are not incidentally overlooked. In the present case many, almost degenerate, inequivalent minima are found.

For physical applications the central question is whether degenerate minima lead to identical classifications of the data. In the case at hand we find that this is to a limited extent the case. A small $<2\%$ fraction of the data exhibits fairly unpredictable re-ordering behavior. To be on the save side, one may combine several network solutions according to a worst case scenario.

**Acknowledgements:** I would like to thank Harrison Prosper and Jeff McDonald for valuable discussions and for supplying me with the data used in this paper. The data, the final parameters and the RC Fortran program are available through e-mail to the author.

# Tables

| 1 | 2 | 3 | 4 | 5 | 6 | 7 |
|---|---|---|---|---|---|---|
| | $\langle \|\|x^{(s)}_{\text{select}} - x^{(s)}_{\text{min}}\|\| \rangle$ | $\langle \|\|x^{(s)}_{\text{min}} - x^{(t)}_{\text{min}}\|\| \rangle$ | $\langle \|\|x^{(s)}_{\text{ran}} - x^{(t)}_{\text{ran}}\|\| \rangle$ | $\langle \|\|Y^{(s)}_{\text{sort}} - Y^{(t)}_{\text{sort}}\|\| \rangle$ | $\langle \|\|Y^{(s)} - Y^{(t)}\|\| \rangle$ | $\langle \|\|Y^{(s)}_{\text{ran}} - Y^{(t)}_{\text{ran}}\|\| \rangle$ |
| $\|\|.\|\|_1$ | 0.0253 (28) | 0.361 (11) | 0.4059 | 0.0217 (17) | 0.0528 (22) | 0.4082 |
| $\|\|.\|\|_2$ | 0.0754 (75) | 0.794 (23) | 0.8427 | 0.0457 (34) | 0.579 (37) | 0.9875 |
| $\|\|.\|\|_3$ | 0.0168 (22) | 0.295 (10) | 1/3 | 0.0180 (15) | 0.0363 (16) | 1/3 |

Table 1: *Average distances between various vectors in parameter and function space.*

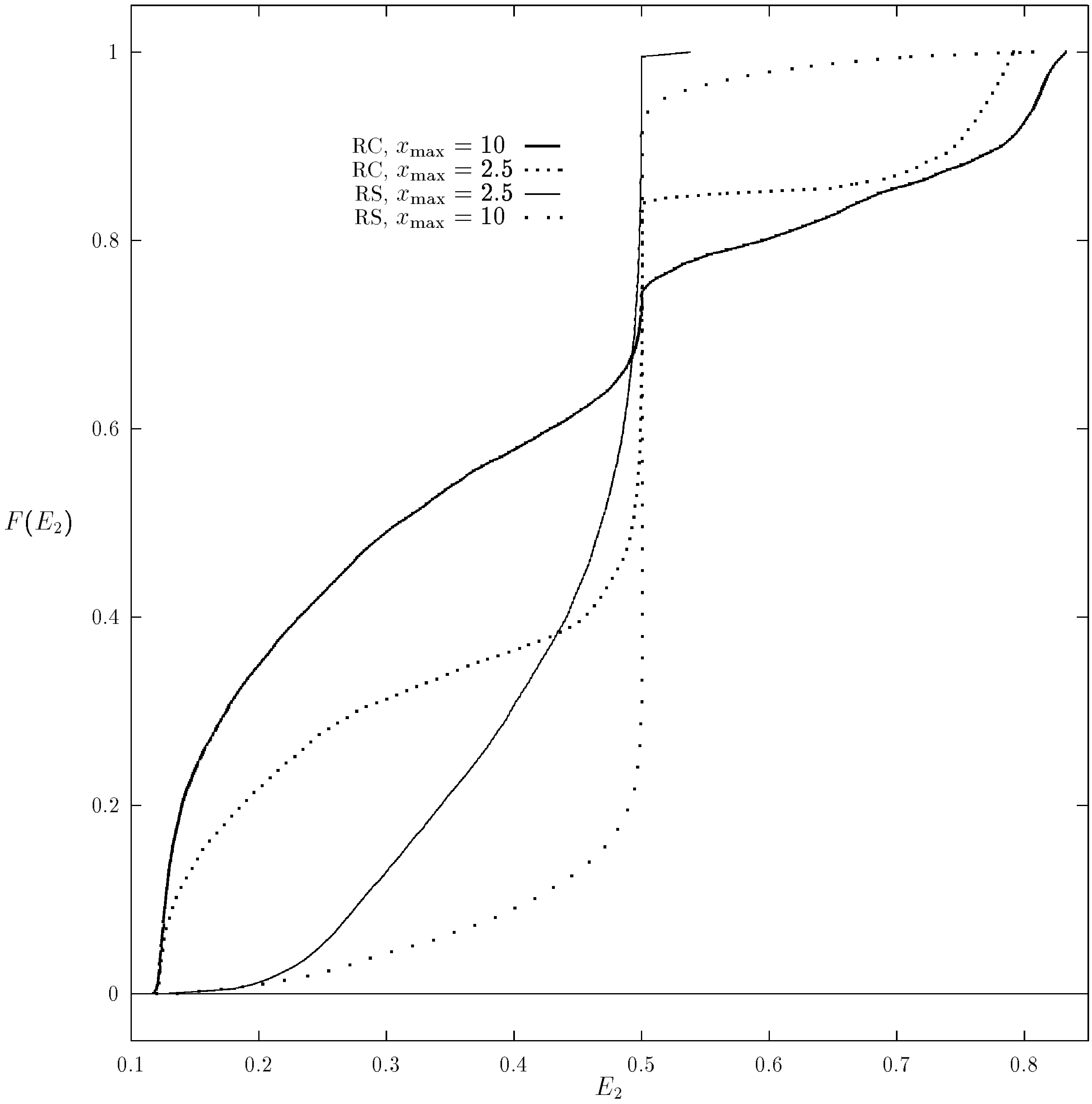

Figure 1: RC and RS distribution functions $F(E_2)$. For $E_2$ small only the RC curves exhibit the desired steep slope.

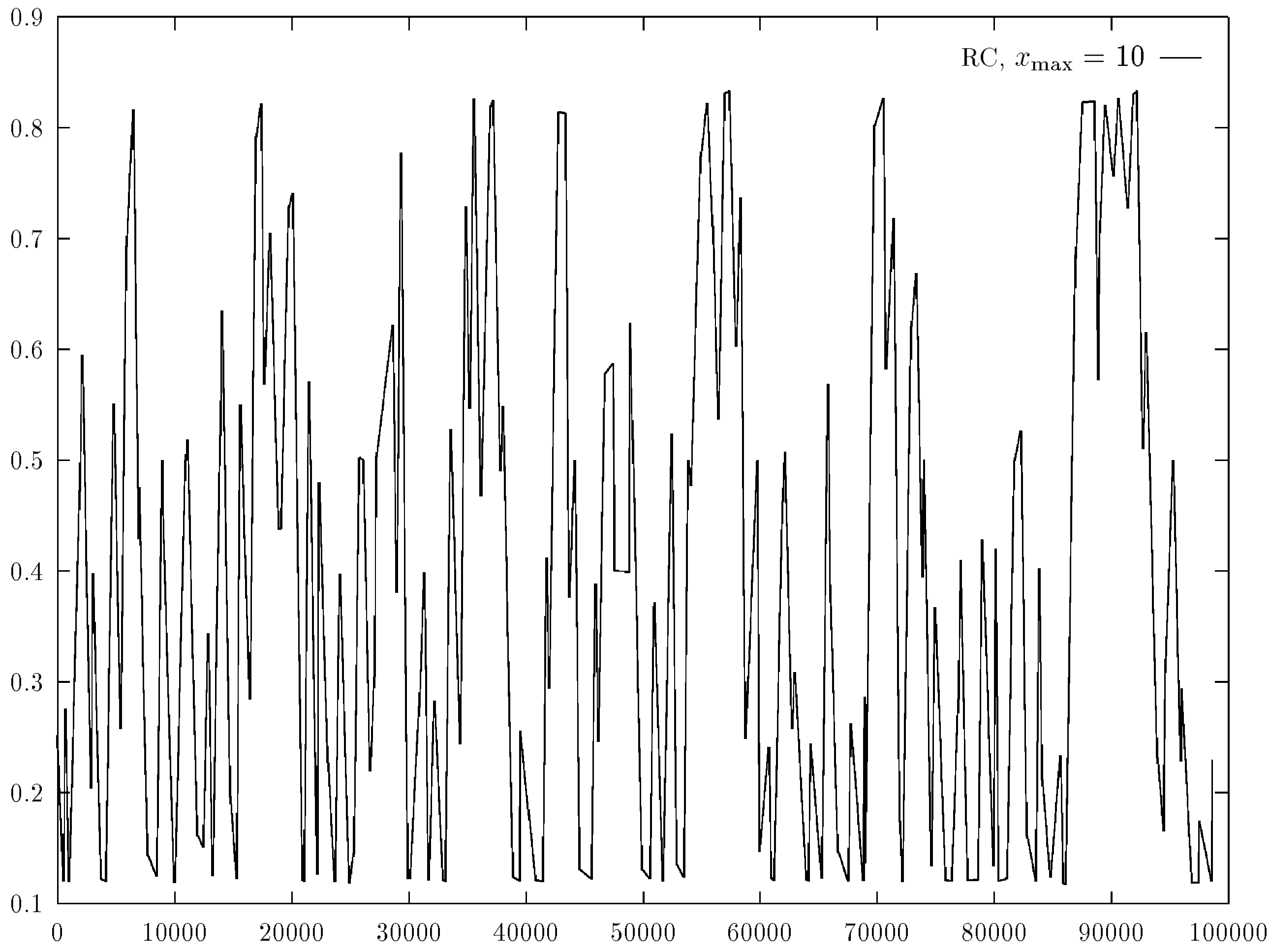

Figure 2: Time series for a RC simulation. For each thousand subsequent data points the minimum and maximum $E_2$ values are connected by straight lines.

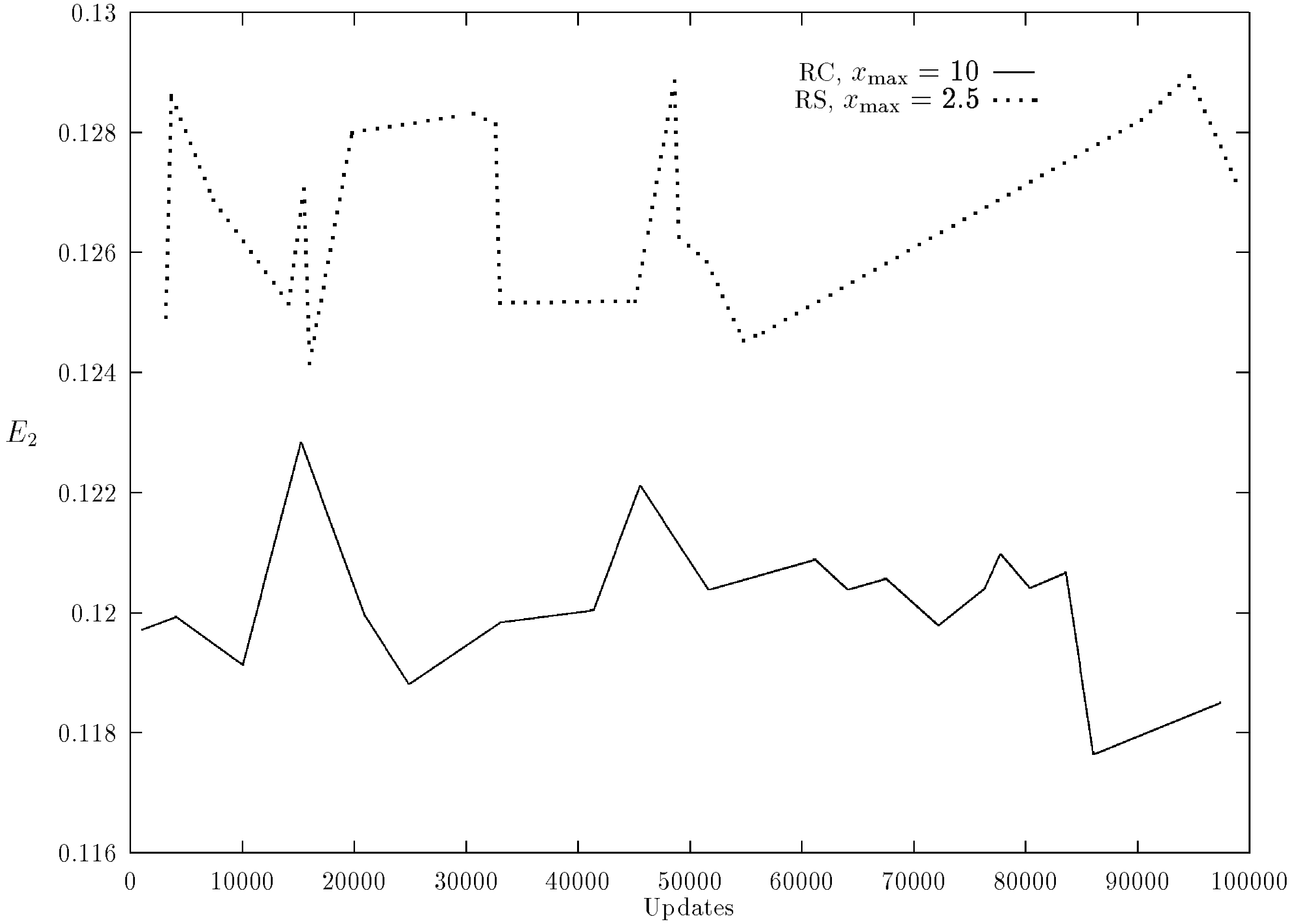

Figure 3: Independent minima reached by 100,000 RC updates in comparison with those from $10^6$ independent RS configurations.

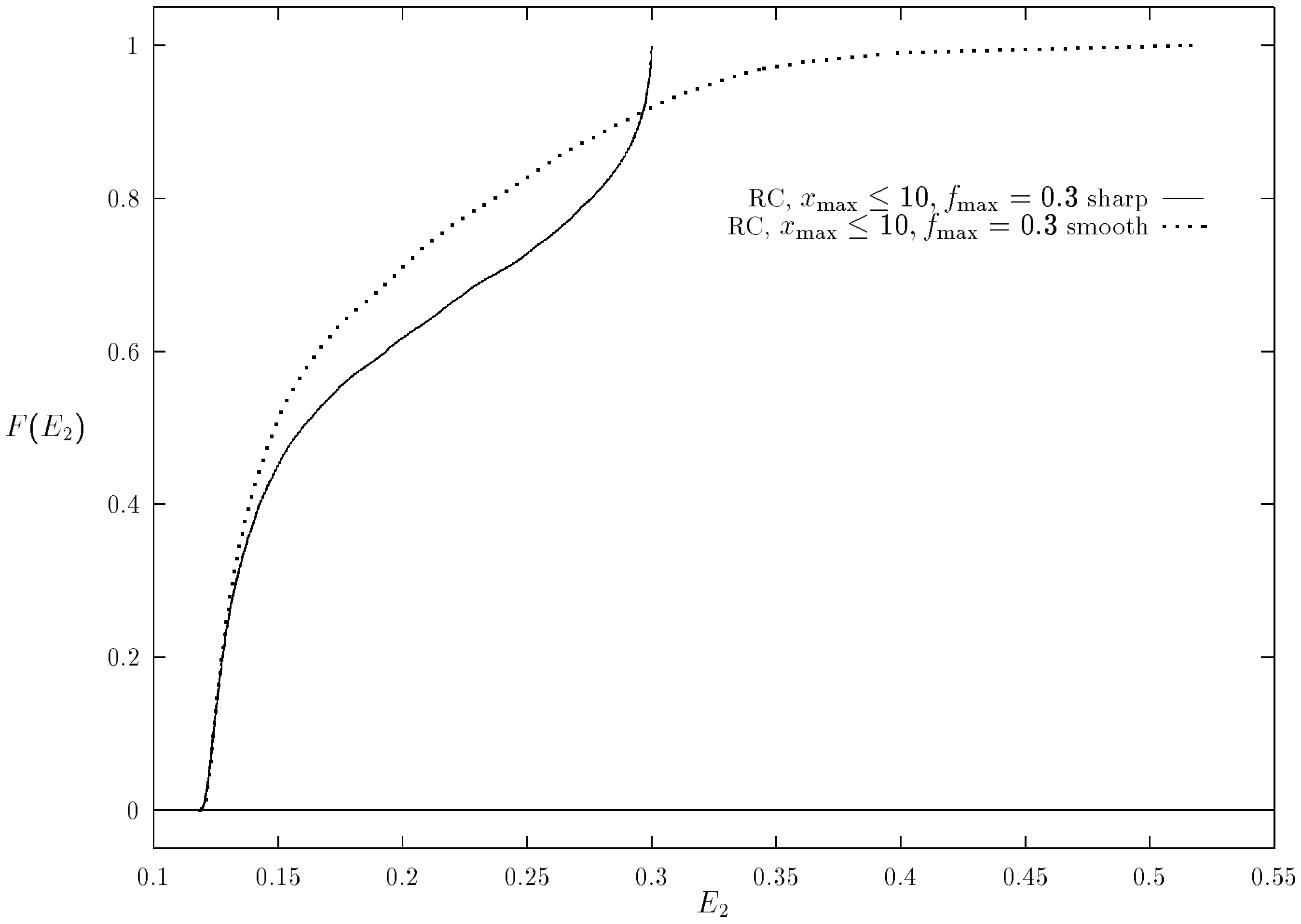

Figure 4: RC distributions functions $F(E_2)$ with a sharp versus a smooth upper bound $f_{\max}$ imposed.

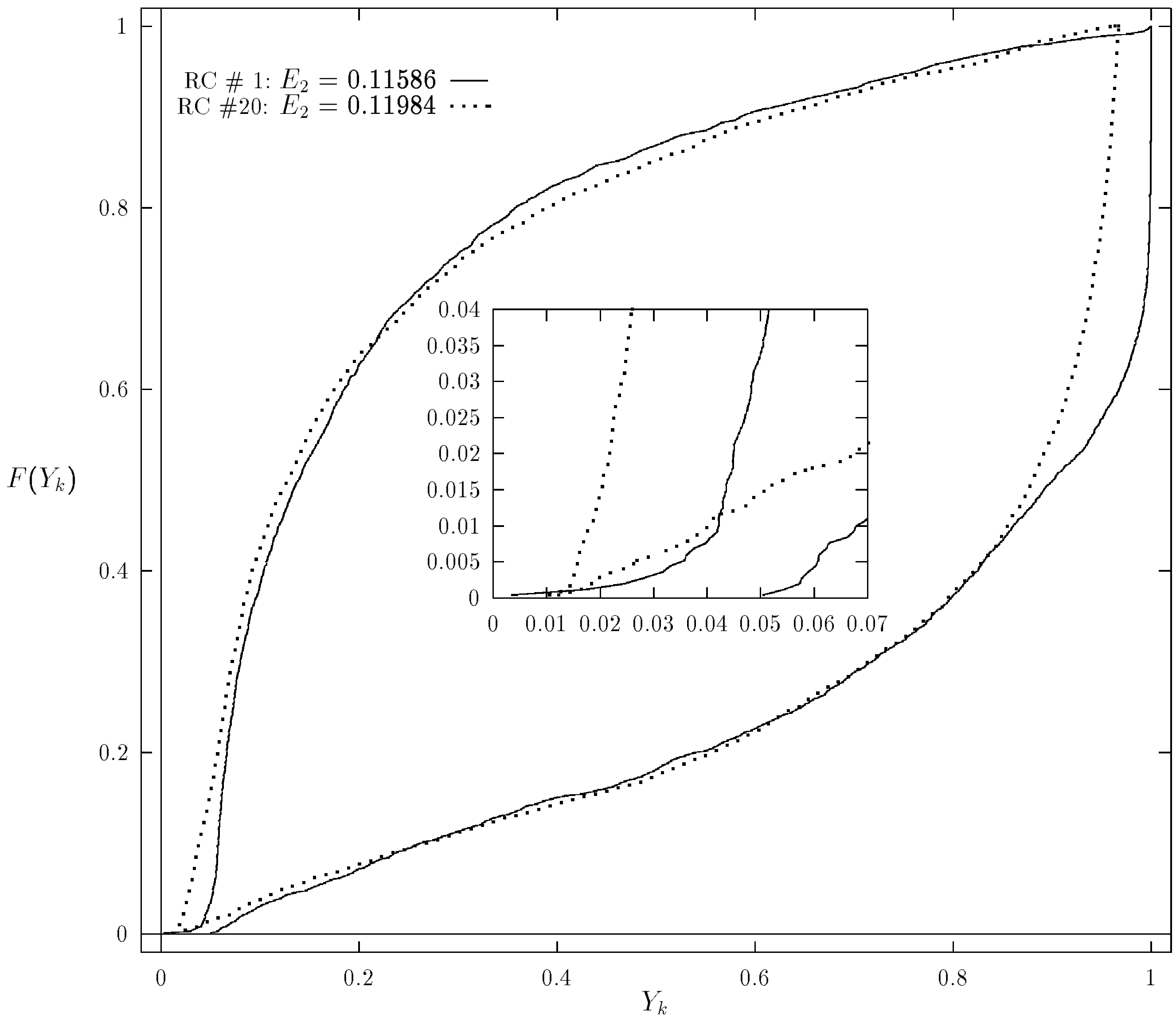

Figure 5: Distributions functions $F(Y_k)$ for event and background training data corresponding to the RC minima # 1 and # 20.

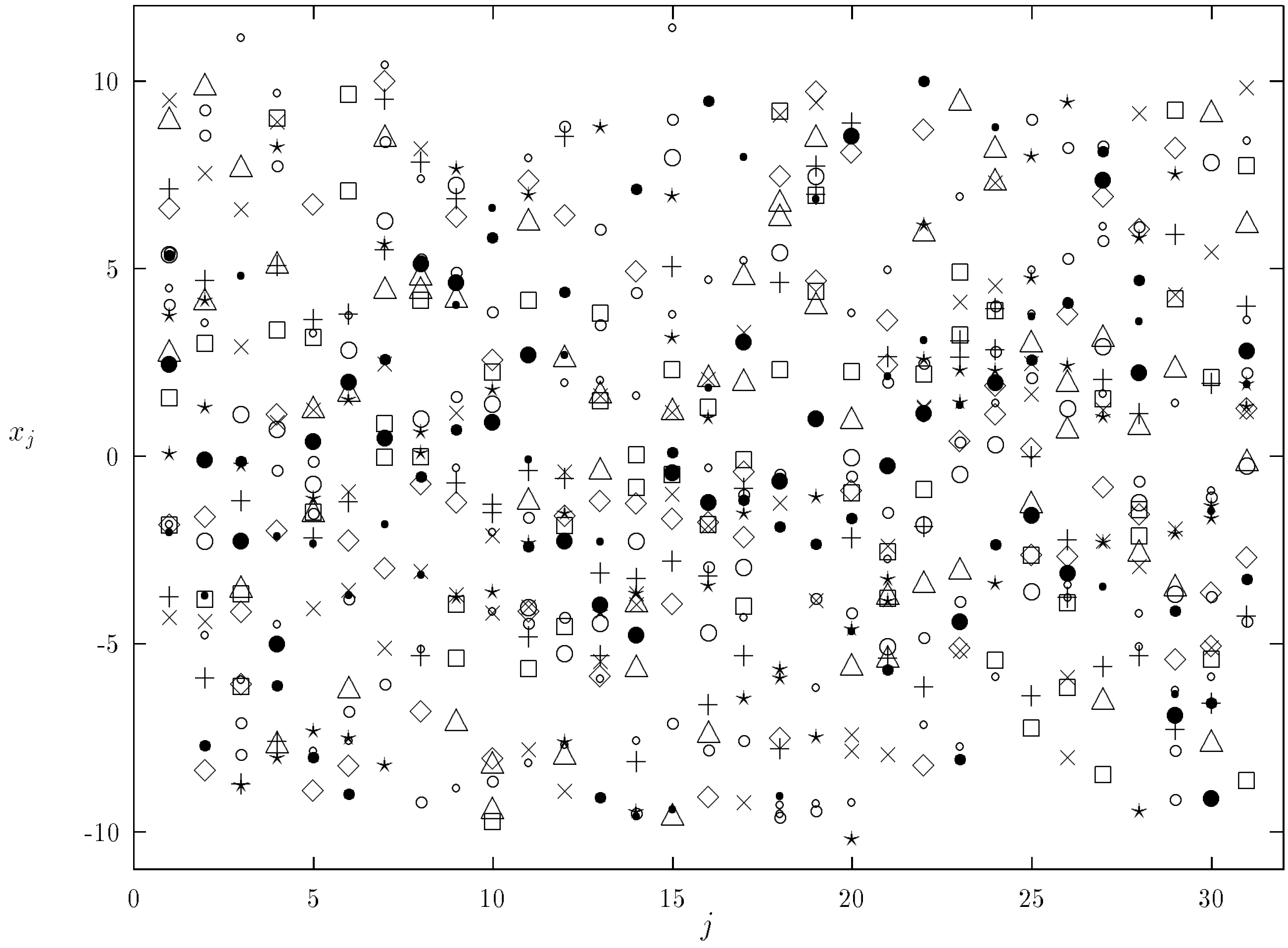

Figure 6: Parameters $x_j$, $j = 1, ..., 31$ for all twenty RC minima.

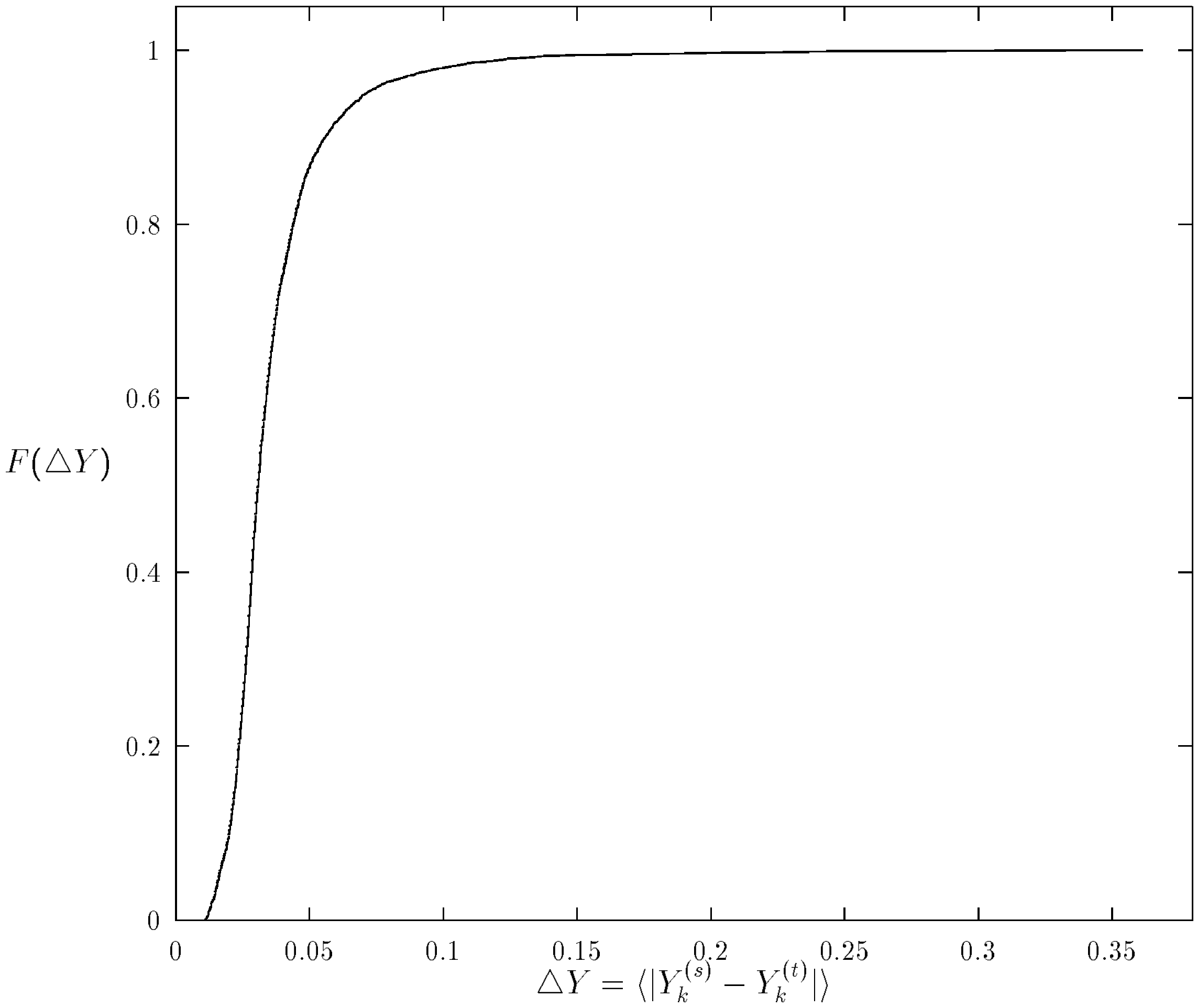

Figure 7: Distribution function of $\langle |Y_k^{(s)} - Y_k^{(t)}| \rangle$ for k=1,...,5000 data points.

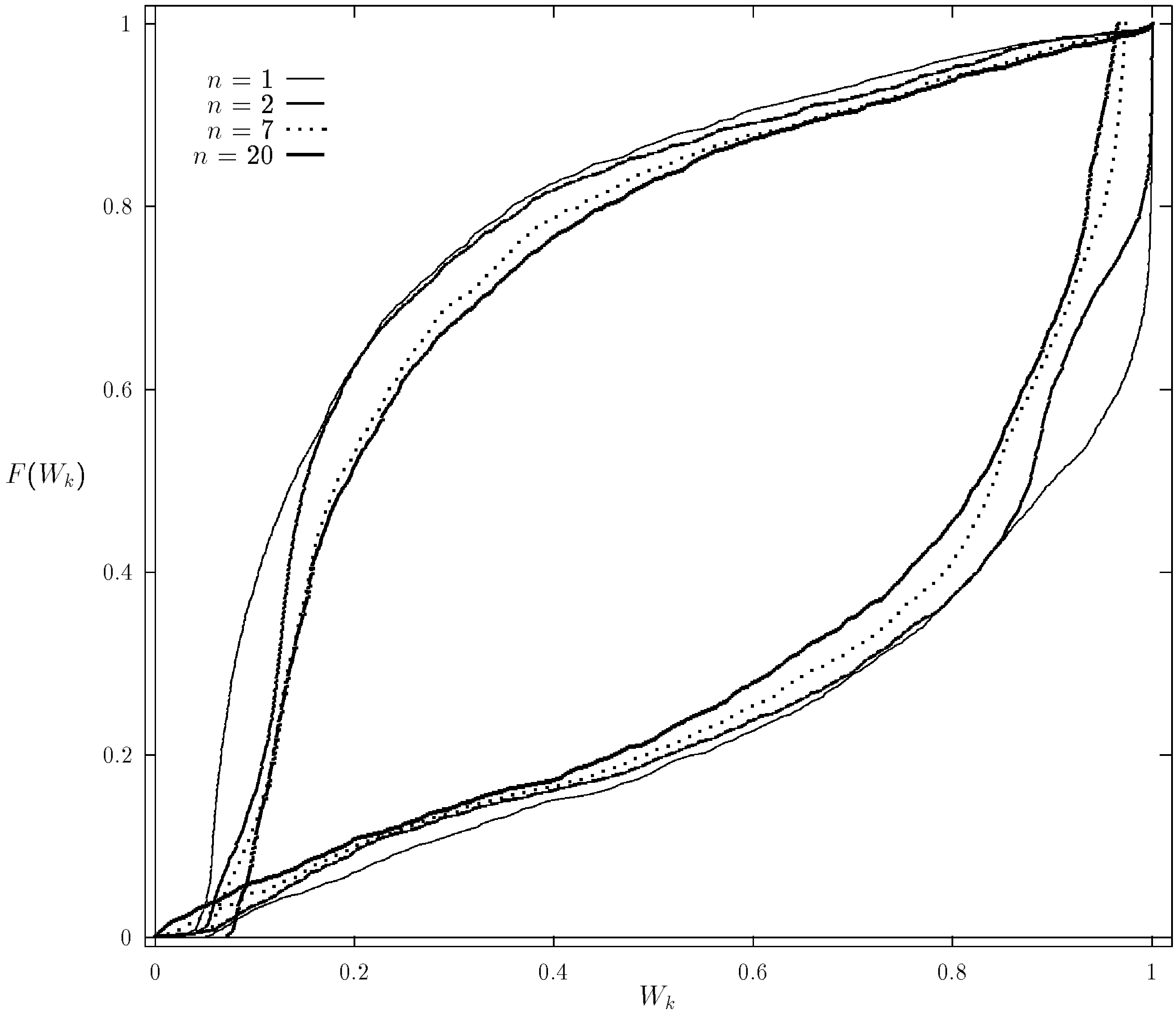

Figure 8: Distributions functions $F(W_k^n)$ for event and background training data obtained by combining RC minima according to the worst case scenario.